\documentclass[11pt]{article}

\usepackage[margin=1in]{geometry}
\usepackage{amsmath,amssymb}
\usepackage{graphicx}
\usepackage{booktabs}
\usepackage[authoryear,round]{natbib}
\usepackage{setspace}
\usepackage{caption}
\usepackage[hidelinks]{hyperref}
\usepackage{xcolor}

\newcommand{\nQual}{256}
\newcommand{\minAB}{300}
\newcommand{\identityErr}{6e-17}
\newcommand{\judgeBA}{0.331}
\newcommand{\wilsonBA}{0.311}
\newcommand{\judgeC}{0.704}
\newcommand{\judgeF}{0.470}
\newcommand{\wilsonC}{0.920}
\newcommand{\wilsonF}{0.338}
\newcommand{\judgeCrank}{43rd}

\newcommand{\rcf}{-0.40}
\newcommand{\rcfLo}{-0.50}
\newcommand{\rcfHi}{-0.29}
\newcommand{\vc}{0.0072}
\newcommand{\vf}{0.0095}
\newcommand{\twocov}{-0.0066}
\newcommand{\vba}{0.0102}
\newcommand{\vind}{0.0167}
\newcommand{\pctred}{39\%}
\newcommand{\nPairs}{19}
\newcommand{\medPairN}{183}
\newcommand{\medC}{0.86}
\newcommand{\minC}{0.81}
\newcommand{\maxC}{0.88}
\newcommand{\medF}{0.44}
\newcommand{\minF}{0.32}
\newcommand{\maxF}{0.51}
\newcommand{\medBA}{0.44}
\newcommand{\minBA}{0.30}
\newcommand{\maxBA}{0.53}
\newcommand{\nBoth}{176}
\newcommand{\rCtwo}{0.82}
\newcommand{\rFtwo}{0.44}
\newcommand{\muc}{1.210}
\newcommand{\muf}{-0.691}
\newcommand{\muclo}{1.163}
\newcommand{\muchi}{1.257}
\newcommand{\muflo}{-0.709}
\newcommand{\mufhi}{-0.674}
\newcommand{\pmuc}{0.770}
\newcommand{\pmuf}{0.334}
\newcommand{\sigc}{0.368}
\newcommand{\sigf}{0.088}
\newcommand{\rhomc}{-0.68}
\newcommand{\sigclo}{0.335}
\newcommand{\sigchi}{0.407}
\newcommand{\sigflo}{0.067}
\newcommand{\sigfhi}{0.109}
\newcommand{\rholo}{-0.85}
\newcommand{\rhohi}{-0.50}
\newcommand{\relc}{0.91}
\newcommand{\relf}{0.37}
\newcommand{\tauc}{0.114}
\newcommand{\tauf}{0.115}

\newcommand{\judgeZc}{-0.342}
\newcommand{\judgeZf}{0.571}
\newcommand{\judgeCs}{0.699}
\newcommand{\judgeFs}{0.381}
\newcommand{\judgeCl}{0.699}
\newcommand{\judgeFl}{0.380}
\newcommand{\mnAbsDc}{0.005}
\newcommand{\mnAbsDf}{0.018}

\newcommand{\nNegro}{114}
\newcommand{\nFourHundred}{13}
\newcommand{\lastFourHundred}{1941}

\newcommand{\sdBAnineteen}{0.037}
\newcommand{\sdBAforties}{0.030}
\newcommand{\sdBAtens}{0.029}
\newcommand{\sdDrop}{19\%}

\newcommand{\rcfPostLo}{-0.38}
\newcommand{\rcfPostHi}{-0.31}

\newcommand{\tvcRatio}{4.6}
\newcommand{\tvfRatio}{0.50}
\newcommand{\nvlfShareNineteen}{36\%}
\newcommand{\nvlfShareTens}{51\%}
\newcommand{\rTalNineteen}{0.09}
\newcommand{\rTalTens}{-0.53}
\newcommand{\talCF}{0.0132}
\newcommand{\talTens}{0.0059}
\newcommand{\talNineteen}{0.0131}

\title{Batting Average as the Product of Two Rates:\\
Skill, Luck, and the Disappearance of the .400 Hitter}
\author{Jim Albert, Emeritus Professor\\[2pt]
\small Department of Mathematics and Statistics, Bowling Green State University}
\date{}

\begin{document}
\maketitle

\begin{abstract}
\noindent
Batting average factors exactly as $\mathrm{BA} = c \times f$, where $c = (AB - SO)/AB$ is the rate of avoiding a strikeout and $f = H/(AB - SO)$ is the rate at which non-strikeout at-bats become hits. Using the \nQual{} major-league hitters with at least \minAB{} at-bats in 2025, we show that the two factors behave very differently. Strikeout avoidance is highly repeatable, with a median year-to-year correlation of \medC{} over \nPairs{} consecutive-season pairs from 2004 to 2025. The finishing rate is not: its median correlation is \medF{}, and batting average itself (\medBA{}) is no more repeatable than its noisier factor. A bivariate logistic-normal random-effects model fit to the 2025 season estimates the correlation between the two talents at $\rho = \rhomc{}$ (95\% profile interval \rholo{} to \rhohi{}), far stronger than the raw correlation of \rcf{}, and implies single-season reliabilities of \relc{} for $c$ and \relf{} for $f$. The model also yields a closed-form bivariate shrinkage estimator in which a hitter's strikeout rate informs the estimate of his finishing rate. Applied to decades of American and National League data, the decomposition revisits Gould's explanation for the disappearance of the .400 hitter. Since the dead-ball era, the talent variance of strikeout avoidance has grown roughly \tvcRatio-fold while that of finishing has halved, and the correlation between them has moved from near zero to about \rTalTens{}. That emerging trade-off, rather than a general narrowing of talent, accounts for the reduced spread of modern batting averages.
\end{abstract}

\noindent\textbf{Keywords:} batting average; empirical Bayes; multivariate random effects; reliability; shrinkage; sabermetrics

\section{Introduction}

Batting average is the oldest summary of hitting performance and among the most studied in statistics. It served as the running example in the empirical Bayes work of \citet{EfronMorris1975,EfronMorris1977}, and \citet{Brown2008} used it to compare empirical Bayes and Bayes methods for in-season prediction. These analyses treat batting average as a single binomial proportion whose true value is estimated by shrinking observed averages toward a common mean \citep{Morris1983}.

Batting average is, however, a composite. An at-bat either ends in a strikeout or it does not, and hits can come only from at-bats that do not. It follows that
\begin{equation}
\mathrm{BA} = \frac{H}{AB} = \underbrace{\frac{AB - SO}{AB}}_{c} \times \underbrace{\frac{H}{AB - SO}}_{f}, \label{eq:identity}
\end{equation}
an exact identity for every hitter in every season. We call $c$ the \emph{strikeout-avoidance rate} and $f$ the \emph{finishing rate}. The rate $c$ is related to, but differs from, contact rates computed per swing. The rate $f$ is close to batting average on balls in play (BABIP), except that home runs are included in both its numerator and its denominator.

The decomposition itself is not new. \citet{Bickel2004} wrote batting average as the product of one minus the strikeout rate and the in-play average to explain why .400 has become so hard to reach, noting that strikeout rates had risen sharply over the twentieth century while in-play averages stayed comparatively stable. \citet{Albert2005} took Bickel's decomposition as the starting point for asking whether batting average measures ability or luck. Fitting beta-binomial random-effects models to eight hitting rates from the 2003 season, he found that strikeout, walk and in-play home-run rates are much more ability-driven than batting average, which confounds ``the propensity of striking out and the ability to make a batted ball fall in for a hit.'' The present paper returns to that question with current data, a joint model for the two factors, and the historical record.

The two factors have different characters. Avoiding a strikeout depends mainly on the hitter: pitch recognition, zone control and bat-to-ball skill. What happens after contact depends also on fielders, park dimensions and chance. \citet{McCracken2001} made the corresponding point for pitchers, arguing that they have little control over hits on balls in play, and the same logic motivates the use of BABIP as an indicator of luck for hitters \citep{Tango2007}. \citet{Albert2016} showed that modeling component rates separately and then combining them gives better predictions of batting and pitching measures than shrinking the aggregate directly, a ``separate and aggregate'' strategy. \citet{Albert2006} applied the same talent-versus-luck reasoning to strikeout rates.

This paper uses identity~\eqref{eq:identity} to ask three questions. How much of the spread in batting average within a season comes from each factor (Section~\ref{sec:cross})? Which factor behaves like a stable skill (Section~\ref{sec:repeat})? And what does a joint model for the two rates add to the usual one-rate-at-a-time analysis (Section~\ref{sec:joint})? Section~\ref{sec:gould} applies the decomposition to more than a century of seasons to revisit \citet{Gould1986,Gould1996}'s explanation for the disappearance of the .400 hitter, and \citet{Bickel2004}'s. Section~\ref{sec:discussion} discusses limitations.

\section{Data}\label{sec:data}

All data come from the Lahman baseball database \citep{Lahman2026}, analyzed in R \citep{R2024}. The main sample consists of the \nQual{} hitters with at least \minAB{} at-bats in the 2025 season, combining stints with different teams. For every hitter, $c \times f$ reproduces batting average to within floating-point error (largest discrepancy \identityErr{}). Section~\ref{sec:repeat} uses all seasons from 2004 to 2025 with the same threshold. Section~\ref{sec:gould} uses American, National and Federal League seasons from 1901 to 2019. We exclude the Negro Leagues, recently added to the database, because their strikeout records are incomplete (\nNegro{} qualifying seasons). All code needed to reproduce the results is available from the author.

\section{The cross-section of hitters in 2025}\label{sec:cross}

\subsection{Equal averages, opposite recipes}

Table~\ref{tab:cases} shows two pairs of hitters who reached similar batting averages in very different ways. Aaron Judge batted \judgeBA{} with a strikeout-avoidance rate of \judgeC{}, the \judgeCrank{} lowest among qualifiers, and a finishing rate of \judgeF{}. Jacob Wilson batted \wilsonBA{} with the second-highest strikeout-avoidance rate in the sample (\wilsonC{}) and a finishing rate of only \wilsonF{}.

\begin{table}[t]
\centering\small
\caption{Four 2025 hitters whose batting averages hide very different $(c, f)$ combinations.}
\label{tab:cases}
\begin{tabular}{lrrrrrr}
\toprule
Player & AB & H & SO & BA & $c$ & $f$ \\
\midrule
Aaron Judge & 541 & 179 & 160 & 0.331 & 0.704 & 0.470 \\
Miguel Andujar & 321 & 102 &  49 & 0.318 & 0.847 & 0.375 \\
Jonathan Aranda & 370 & 117 & 107 & 0.316 & 0.711 & 0.445 \\
Jacob Wilson & 486 & 151 &  39 & 0.311 & 0.920 & 0.338
 \\
\bottomrule
\end{tabular}
\end{table}

\begin{figure}[t]
\centering
\includegraphics[width=0.85\textwidth]{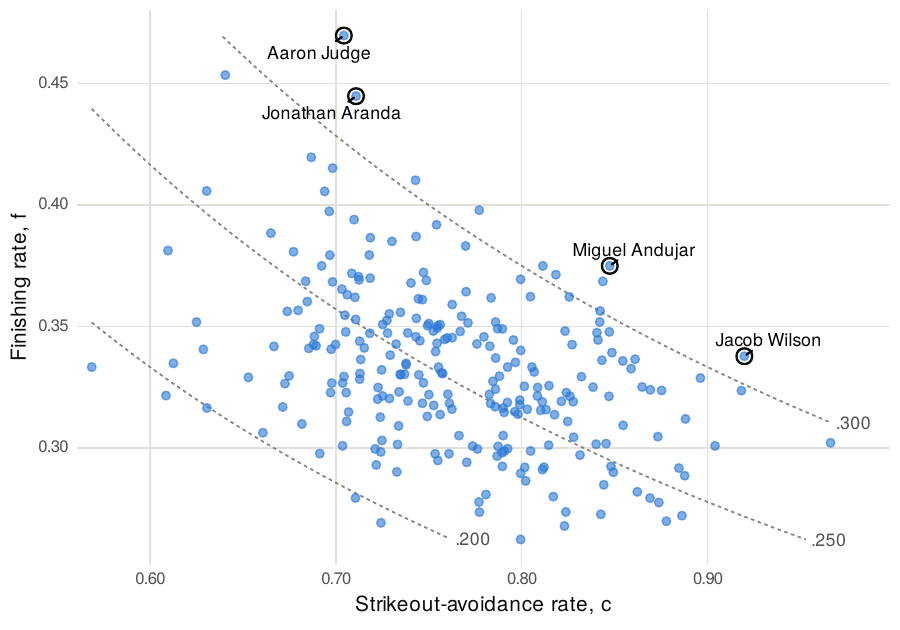}
\caption{Strikeout-avoidance rate $c$ and finishing rate $f$ for the \nQual{} hitters with at least \minAB{} at-bats in 2025. Dashed curves are lines of constant batting average.}
\label{fig:scatter}
\end{figure}

Figure~\ref{fig:scatter} places all \nQual{} qualifiers in the $(c, f)$ plane. The two rates are negatively correlated ($r = \rcf{}$, 95\% interval \rcfLo{} to \rcfHi{}): hitters who rarely strike out tend to do less with their contact, and the best finishers strike out more. Lines of constant batting average are hyperbolas $f = \mathrm{BA}/c$, and the cloud of hitters lies partly along them.

\subsection{A variance decomposition}

Taking logarithms of \eqref{eq:identity} gives $\log \mathrm{BA} = \log c + \log f$, so the cross-sectional variance decomposes exactly as
\begin{equation}
\mathrm{Var}(\log \mathrm{BA}) = \mathrm{Var}(\log c) + \mathrm{Var}(\log f) + 2\,\mathrm{Cov}(\log c, \log f). \label{eq:vardecomp}
\end{equation}
In 2025 the three terms are \vc{}, \vf{} and \twocov{}, summing to $\mathrm{Var}(\log \mathrm{BA}) = \vba{}$. The two factors contribute similar amounts of variance. Because they are negatively correlated, however, the observed spread in batting average is about \pctred{} smaller than it would be if $c$ and $f$ varied independently (\vind{}).

\section{Which rate is a skill?}\label{sec:repeat}

A cross-sectional contribution to variance does not by itself indicate a skill; a skill should persist from one season to the next. For every pair of consecutive seasons from 2004 to 2025, we correlate each rate for the hitters who qualified in both seasons. The 2020 season, in which almost no one reached \minAB{} at-bats, drops out, leaving \nPairs{} pairs with a median of \medPairN{} hitters each.

\begin{figure}[t]
\centering
\includegraphics[width=0.9\textwidth]{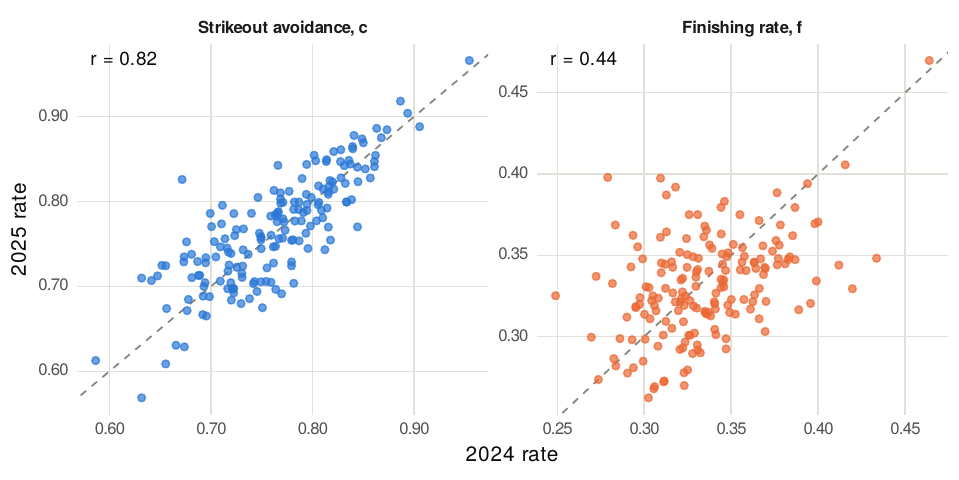}
\caption{2024 and 2025 rates for the \nBoth{} hitters with at least \minAB{} at-bats in both seasons. The dashed line is $y = x$.}
\label{fig:repeat}
\end{figure}

\begin{figure}[t]
\centering
\includegraphics[width=0.65\textwidth]{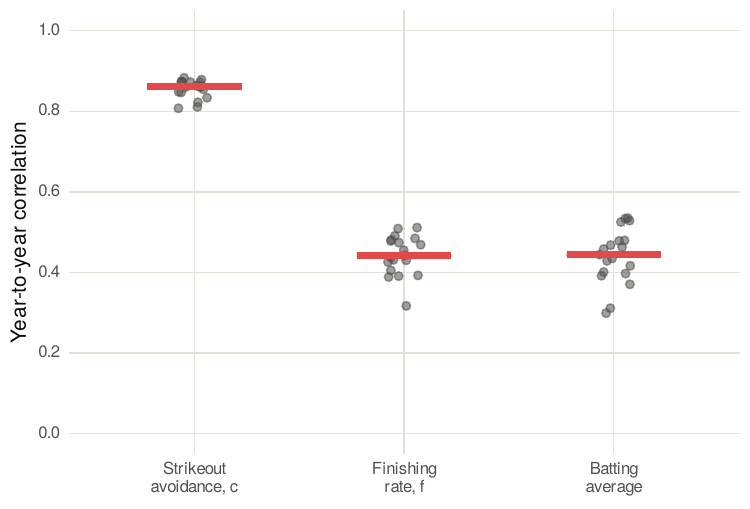}
\caption{Year-to-year correlations for each of the \nPairs{} consecutive-season pairs from 2004 to 2025. Horizontal bars mark the medians.}
\label{fig:stability}
\end{figure}

Figure~\ref{fig:repeat} shows the 2024--2025 pair (\nBoth{} hitters). Strikeout avoidance repeats closely ($r = \rCtwo{}$), and finishing rate much less so ($r = \rFtwo{}$). Figure~\ref{fig:stability} shows that this pattern holds in every pair. The year-to-year correlation for $c$ ranges from \minC{} to \maxC{} (median \medC{}). For $f$ it ranges from \minF{} to \maxF{} (median \medF{}), and for batting average from \minBA{} to \maxBA{} (median \medBA{}). The repeatability of batting average matches that of its less reliable factor rather than falling between the two.

\section{A joint model for the two rates}\label{sec:joint}

\subsection{Model}

Hierarchical models for batting data usually treat one rate at a time. Because $c$ and $f$ are correlated, we instead model them jointly. For hitter $i$,
\begin{align}
AB_i - SO_i &\sim \mathrm{Binomial}(AB_i, c_i), & H_i &\sim \mathrm{Binomial}(AB_i - SO_i, f_i), \nonumber\\
\mathrm{logit}(c_i) &= \mu_c + u_i, & \mathrm{logit}(f_i) &= \mu_f + v_i, \label{eq:model}\\
(u_i, v_i)^\top &\sim \mathrm{N}_2(\mathbf{0}, \Sigma), & \Sigma &= \begin{pmatrix} \sigma_c^2 & \rho\sigma_c\sigma_f \\ \rho\sigma_c\sigma_f & \sigma_f^2 \end{pmatrix}. \nonumber
\end{align}
Here $\sigma_c$ and $\sigma_f$ measure the spread of true talent on the logit scale, and $\rho$ is the correlation between the two talents with sampling noise removed. Given the talents, the two binomial outcomes are independent: whether an at-bat ends in a strikeout and whether a non-strikeout at-bat becomes a hit are determined by disjoint events. We fit \eqref{eq:model} by maximum likelihood with the Laplace approximation in \texttt{lme4} \citep{Bates2015}, stacking the two outcomes and specifying an unstructured covariance for the two random intercepts, and compute profile-likelihood intervals for the variance components.

\subsection{Estimates}

Table~\ref{tab:params} gives the estimates. The means correspond to a league strikeout-avoidance rate of \pmuc{} and finishing rate of \pmuf{}. The estimated talent correlation, $\hat\rho = \rhomc{}$, is much stronger than the raw correlation of \rcf{}. The raw correlation is attenuated because sampling noise inflates the variance of each observed rate without changing their covariance, the classical attenuation described by \citet{Spearman1904}.

\begin{table}[t]
\centering\small
\caption{Maximum likelihood estimates for model \eqref{eq:model}, 2025 season (logit scale). Intervals are Wald intervals for the means and profile-likelihood intervals for the variance components.}
\label{tab:params}
\begin{tabular}{llr@{\hspace{1.5em}}l}
\toprule
Parameter & Interpretation & Estimate & 95\% interval \\
\midrule
$\mu_c$ & league strikeout avoidance & \muc{} & (\muclo{}, \muchi{}) \\
$\mu_f$ & league finishing rate & \muf{} & (\muflo{}, \mufhi{}) \\
$\sigma_c$ & talent SD, strikeout avoidance & \sigc{} & (\sigclo{}, \sigchi{}) \\
$\sigma_f$ & talent SD, finishing & \sigf{} & (\sigflo{}, \sigfhi{}) \\
$\rho$ & talent correlation & \rhomc{} & (\rholo{}, \rhohi{}) \\
\bottomrule
\end{tabular}
\end{table}

\subsection{Single-season reliability}

By the delta method, the sampling variance of hitter $i$'s observed logit rates is approximately $\tau^2_{i,c} = 1/\{AB_i\, \hat c_i(1-\hat c_i)\}$ and $\tau^2_{i,f} = 1/\{(AB_i - SO_i)\, \hat f_i(1-\hat f_i)\}$. The ratio
\begin{equation}
w_{i,c} = \frac{\sigma_c^2}{\sigma_c^2 + \tau_{i,c}^2}, \qquad w_{i,f} = \frac{\sigma_f^2}{\sigma_f^2 + \tau_{i,f}^2}
\end{equation}
is the fraction of the observed between-hitter variation that reflects talent, the reliability of classical test theory \citep{LordNovick1968}. The noise SDs are nearly equal for the two rates (mean $\tau$ of \tauc{} and \tauf{}), but the talent SDs are not. Averaged over the \nQual{} hitters, the reliability is \relc{} for strikeout avoidance and \relf{} for finishing. These values come from a single season, yet they agree with the multi-season correlations of Section~\ref{sec:repeat} (\medC{} and \medF{}). This is expected, since reliability and test--retest correlation estimate the same quantity when talent is stable.

\subsection{Bivariate shrinkage}

Let $z_i = (\mathrm{logit}\,\hat c_i - \mu_c,\ \mathrm{logit}\,\hat f_i - \mu_f)^\top$ and $T_i = \mathrm{diag}(\tau^2_{i,c}, \tau^2_{i,f})$. Under a normal approximation to the binomial likelihood on the logit scale, the posterior mean of hitter $i$'s talents is
\begin{equation}
\hat b_i = \Sigma\,(\Sigma + T_i)^{-1} z_i. \label{eq:shrink}
\end{equation}
When $\rho = 0$, this reduces to shrinking each rate separately by its own reliability, the familiar James--Stein form \citep{EfronMorris1975}. When $\rho \neq 0$, the off-diagonal terms of $\Sigma(\Sigma + T_i)^{-1}$ let a hitter's strikeout avoidance inform the estimate of his finishing talent, and vice versa.

\begin{figure}[t]
\centering
\includegraphics[width=0.85\textwidth]{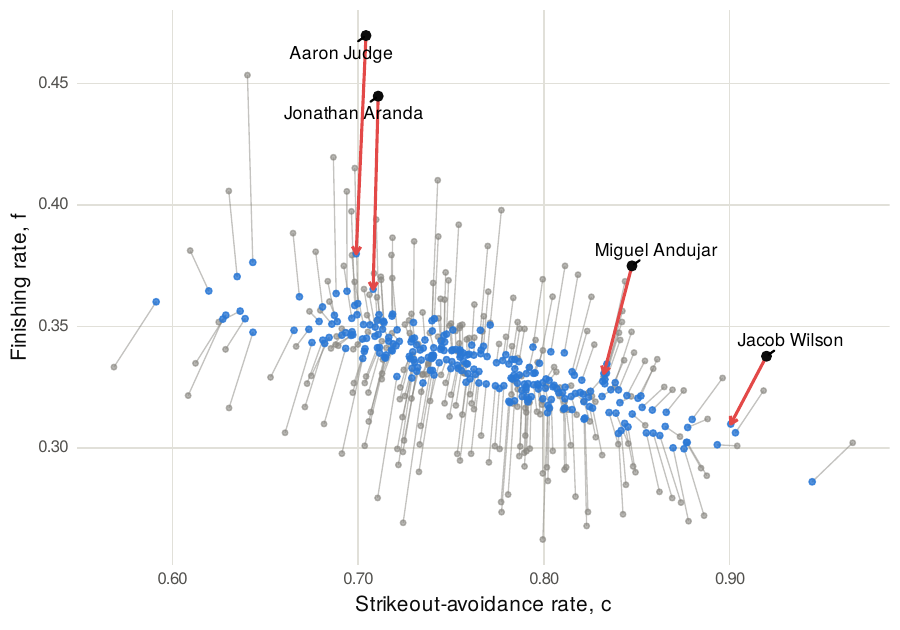}
\caption{Observed rates (gray) and joint-model talent estimates (blue) for 2025 qualifiers. Arrows point from observed to estimated. The four hitters of Table~\ref{tab:cases} are highlighted.}
\label{fig:shrink}
\end{figure}

For Judge, $z = (\judgeZc{}, \judgeZf{})$. Equation~\eqref{eq:shrink} gives estimated talents of \judgeCs{} for $c$ and \judgeFs{} for $f$. The conditional modes from the fitted model, which use the full binomial likelihood rather than a normal approximation, are nearly identical (\judgeCl{} and \judgeFl{}). Judge's finishing estimate stays well above the league mean of \pmuf{} partly because his low strikeout-avoidance rate, through the negative correlation, points to a higher finishing talent. Figure~\ref{fig:shrink} and Table~\ref{tab:shrink} show the shrinkage for all hitters. The mean absolute adjustment is \mnAbsDc{} for $c$ but \mnAbsDf{} for $f$. The finishing rates move the most, and they move toward the negative trend of the joint distribution.

\begin{table}[t]
\centering\small
\caption{Observed rates and joint-model estimates (conditional modes) for the four hitters of Table~\ref{tab:cases}.}
\label{tab:shrink}
\begin{tabular}{lrrrr}
\toprule
Player & $\hat c$ observed & $c$ estimated & $\hat f$ observed & $f$ estimated \\
\midrule
Aaron Judge & 0.704 & 0.699 & 0.470 & 0.380 \\
Miguel Andujar & 0.847 & 0.832 & 0.375 & 0.331 \\
Jonathan Aranda & 0.711 & 0.708 & 0.445 & 0.365 \\
Jacob Wilson & 0.920 & 0.901 & 0.338 & 0.310
 \\
\bottomrule
\end{tabular}
\end{table}

\section{A historical perspective on the .400 hitter}\label{sec:gould}

\begin{figure}[t]
\centering
\includegraphics[width=\textwidth]{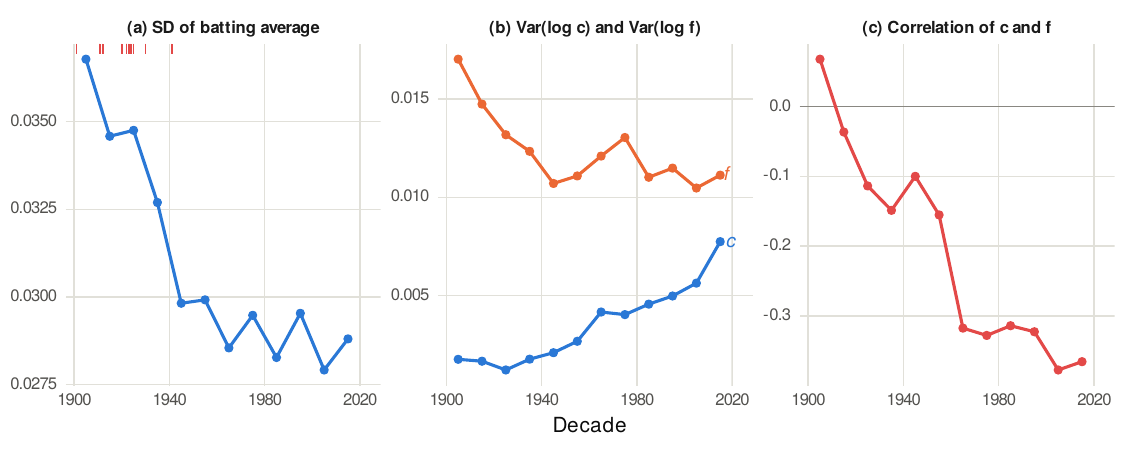}
\caption{American, National and Federal League hitters with at least \minAB{} at-bats, by decade, 1900s--2010s. (a) Standard deviation of batting average; ticks mark the \nFourHundred{} .400 seasons. (b) Observed variances of $\log c$ and $\log f$. (c) Correlation between $c$ and $f$.}
\label{fig:history}
\end{figure}

\citet{Gould1986,Gould1996} argued that the .400 hitter disappeared not because hitters became worse but because the variation in batting average shrank. As play improved and the talent pool deepened, the distance between average and exceptional hitters narrowed, trimming both tails of the distribution. Among American and National League hitters with at least \minAB{} at-bats, there have been \nFourHundred{} seasons of .400 or better since 1901, the last in \lastFourHundred{}. The standard deviation of batting average among qualifiers fell from \sdBAnineteen{} in the 1900s to \sdBAforties{} in the 1940s, a decline of about \sdDrop{}, and has been nearly constant since (\sdBAtens{} in the 2010s; Figure~\ref{fig:history}a).

Gould worked with batting average alone. \citet{Bickel2004} tracked the league-wide levels of the two factors, the rise in strikeouts and the stability of in-play average. Decomposition~\eqref{eq:vardecomp} adds their spreads and their correlation across hitters, showing how the factors changed (Table~\ref{tab:decades}, Figure~\ref{fig:history}). Observed decade-level variances include binomial sampling noise, which matters most for $f$: noise accounts for about \nvlfShareNineteen{} of the observed variance of $\log f$ in the 1900s and \nvlfShareTens{} in the 2010s. We therefore also report talent variances, obtained by subtracting the average delta-method sampling variance, $(1-p)/(np)$ for a rate $p$ based on $n$ trials. Because the sampling errors in $c$ and $f$ are independent, the covariance needs no correction.

\begin{table}[t]
\centering\footnotesize
\setlength{\tabcolsep}{4pt}
\caption{Decade summaries for qualifying hitters. Columns 3--8 are observed quantities. The last two columns remove binomial sampling noise from $\mathrm{Var}(\log f)$ and from the correlation.}
\label{tab:decades}
\begin{tabular}{lrrrrrrrrr}
\toprule
Decade & $n$ & SD(BA) & Var$(\log c)$ & Var$(\log f)$ & $2\,$Cov & Var$(\log \mathrm{BA})$ & $r(c,f)$ & Talent Var$(\log f)$ & Talent $r$ \\
\midrule
1900s & 1105 & 0.037 & 0.0018 & 0.0170 & 0.0007 & 0.0195 & 0.07 & 0.0109 & 0.09 \\
1910s & 1276 & 0.035 & 0.0017 & 0.0147 & -0.0004 & 0.0160 & -0.04 & 0.0088 & -0.06 \\
1920s & 1241 & 0.035 & 0.0012 & 0.0132 & -0.0009 & 0.0136 & -0.11 & 0.0081 & -0.15 \\
1930s & 1296 & 0.033 & 0.0018 & 0.0123 & -0.0013 & 0.0128 & -0.15 & 0.0071 & -0.20 \\
1940s & 1211 & 0.030 & 0.0021 & 0.0107 & -0.0010 & 0.0119 & -0.10 & 0.0050 & -0.16 \\
1950s & 1227 & 0.030 & 0.0027 & 0.0111 & -0.0017 & 0.0121 & -0.16 & 0.0053 & -0.23 \\
1960s & 1581 & 0.029 & 0.0042 & 0.0121 & -0.0044 & 0.0118 & -0.32 & 0.0062 & -0.46 \\
1970s & 2038 & 0.029 & 0.0040 & 0.0130 & -0.0046 & 0.0124 & -0.33 & 0.0072 & -0.45 \\
1980s & 2114 & 0.028 & 0.0046 & 0.0110 & -0.0044 & 0.0112 & -0.31 & 0.0052 & -0.47 \\
1990s & 2229 & 0.030 & 0.0050 & 0.0115 & -0.0048 & 0.0116 & -0.32 & 0.0059 & -0.47 \\
2000s & 2545 & 0.028 & 0.0056 & 0.0105 & -0.0057 & 0.0104 & -0.38 & 0.0051 & -0.56 \\
2010s & 2494 & 0.029 & 0.0077 & 0.0111 & -0.0066 & 0.0122 & -0.37 & 0.0054 & -0.53
 \\
\bottomrule
\end{tabular}
\end{table}

Three trends emerge. First, strikeout avoidance did not homogenize. Its talent variance on the log scale grew by a factor of about \tvcRatio{} from the 1900s to the 2010s, as the game separated hitters who rarely strike out from those who accept strikeouts in exchange for power. Second, the talent variance of finishing roughly halved (ratio \tvfRatio{}), consistent with Gould's picture of a converging talent pool. Third, and most striking, the two rates were essentially uncorrelated in the dead-ball era (talent correlation \rTalNineteen{}), while since the 1960s the observed correlation has stayed between \rcfPostHi{} and \rcfPostLo{} (talent correlation \rTalTens{} in the 2010s). The power--contact trade-off is a twentieth-century development rather than a permanent feature of hitting.

The decomposition allows a counterfactual comparison. On the talent scale, the variance of $\log \mathrm{BA}$ fell from \talNineteen{} in the 1900s to \talTens{} in the 2010s. Combining the 2010s talent variances of $\log c$ and $\log f$ with the 1900s covariance gives \talCF{}, essentially the dead-ball level. The growth in the spread of strikeout avoidance and the narrowing of finishing talent roughly offset, and the reduced spread of modern batting averages is due almost entirely to the stronger negative covariance. In this sense Gould's account is incomplete: a .400 season requires excellence in both $c$ and $f$ at once, and the modern game rarely allows that combination.

\section{Discussion}\label{sec:discussion}

The identity $\mathrm{BA} = c \times f$ separates batting average into a component that behaves like a skill and one that behaves largely like noise. Because it is an exact identity, the variance decomposition and the counterfactual comparison involve no modeling assumptions. The joint model shows that the correlation between the two talents is much stronger than the raw correlation suggests. It also gives a bivariate shrinkage estimator that uses each rate to inform the other, extending the component approach of \citet{Albert2016} by modeling the dependence between components. The conclusion of \citet{Albert2005}, that batting average is a comparatively poor measure of hitting ability, holds two decades later with a different model and data source, and the decomposition now shows which of its two parts is responsible.

Several limitations should be noted. First, $f$ is not a pure luck measure. It includes home runs and reflects real differences in how hard and at what angle hitters strike the ball, which is why its reliability is modest rather than zero. Batted-ball tracking data could separate the contact-quality part of $f$ from the fielding and chance part. Second, at-bats exclude walks, hit-by-pitches and sacrifices, so $c$ and $f$ describe the at-bat outcomes that enter batting average rather than all plate appearances. Third, the \minAB{}-at-bat threshold selects hitters whose teams kept playing them, which may truncate the lower tail of both rates. Fourth, the model treats each season separately. A model with player effects shared across seasons, or a fully Bayesian fit with priors on $\Sigma$ \citep{GelmanHill2007}, would give sharper talent estimates and full posterior uncertainty for $\rho$. Finally, the historical analysis describes when the trade-off emerged, not why. Candidate explanations include the rise of hard-throwing relief pitching, changes in defensive positioning, and hitters' deliberate exchange of contact for power; separating these would require data beyond season totals.

\section*{Acknowledgments and use of AI tools}

This paper grew out of a series of analyses, carried out in August and September 2026, that revisited the question posed in \citet{Albert2005} with current data. The analyses and this manuscript were produced in collaboration with Claude (Anthropic), a large language model. The author posed the research questions, chose which directions to pursue, and reviewed the results at each stage. Claude wrote the R code, carried out the computations, drafted the text, and assembled the references. The author checked the results and the text and corrected errors found in review, including one that had overstated the shrinkage shown in Figure~\ref{fig:shrink}. Claude is not listed as an author because it cannot take responsibility for the work; the author takes full responsibility for the content, including any remaining errors. A single R script reproduces every number, table and figure and is available from the author.

\bibliographystyle{apalike}
\bibliography{refs}

\begin{thebibliography}{}

\bibitem[Albert, 2005]{Albert2005}
Albert, J. (2005).
\newblock Does a baseball hitter's batting average measure ability or luck?
\newblock {\em STATS: The Magazine for Students of Statistics}, (44):3--6.
\newblock Fall 2005/Winter 2006 issue.

\bibitem[Albert, 2006]{Albert2006}
Albert, J. (2006).
\newblock Pitching statistics, talent and luck, and the best strikeout seasons
  of all-time.
\newblock {\em Journal of Quantitative Analysis in Sports}, 2(1).

\bibitem[Albert, 2016]{Albert2016}
Albert, J. (2016).
\newblock Improved component predictions of batting and pitching measures.
\newblock {\em Journal of Quantitative Analysis in Sports}, 12(2):73--85.

\bibitem[Bates et~al., 2015]{Bates2015}
Bates, D., M{\"a}chler, M., Bolker, B., and Walker, S. (2015).
\newblock Fitting linear mixed-effects models using {lme4}.
\newblock {\em Journal of Statistical Software}, 67(1):1--48.

\bibitem[Bickel, 2004]{Bickel2004}
Bickel, J.~E. (2004).
\newblock Why it's so hard to hit .400: New insights into an old statistic.
\newblock {\em The Baseball Research Journal}, 32:15--21.

\bibitem[Brown, 2008]{Brown2008}
Brown, L.~D. (2008).
\newblock In-season prediction of batting averages: A field test of empirical
  {Bayes} and {Bayes} methodologies.
\newblock {\em The Annals of Applied Statistics}, 2(1):113--152.

\bibitem[Efron and Morris, 1975]{EfronMorris1975}
Efron, B. and Morris, C. (1975).
\newblock Data analysis using {Stein's} estimator and its generalizations.
\newblock {\em Journal of the American Statistical Association},
  70(350):311--319.

\bibitem[Efron and Morris, 1977]{EfronMorris1977}
Efron, B. and Morris, C. (1977).
\newblock {Stein's} paradox in statistics.
\newblock {\em Scientific American}, 236(5):119--127.

\bibitem[Friendly et~al., 2026]{Lahman2026}
Friendly, M., Dalzell, C., Monkman, M., and Murphy, D. (2026).
\newblock {\em {Lahman}: Sean `{Lahman}' Baseball Database}.
\newblock R package version 14.0-0.

\bibitem[Gelman and Hill, 2007]{GelmanHill2007}
Gelman, A. and Hill, J. (2007).
\newblock {\em Data Analysis Using Regression and Multilevel/Hierarchical
  Models}.
\newblock Cambridge University Press, Cambridge.

\bibitem[Gould, 1986]{Gould1986}
Gould, S.~J. (1986).
\newblock Entropic homogeneity isn't why no one hits .400 any more.
\newblock {\em Discover}.

\bibitem[Gould, 1996]{Gould1996}
Gould, S.~J. (1996).
\newblock {\em Full House: The Spread of Excellence from {Plato} to {Darwin}}.
\newblock Harmony Books, New York.

\bibitem[Lord and Novick, 1968]{LordNovick1968}
Lord, F.~M. and Novick, M.~R. (1968).
\newblock {\em Statistical Theories of Mental Test Scores}.
\newblock Addison-Wesley, Reading, MA.

\bibitem[McCracken, 2001]{McCracken2001}
McCracken, V. (2001).
\newblock Pitching and defense: How much control do hurlers have?
\newblock Baseball Prospectus.
\newblock January 23, 2001.

\bibitem[Morris, 1983]{Morris1983}
Morris, C.~N. (1983).
\newblock Parametric empirical {Bayes} inference: Theory and applications.
\newblock {\em Journal of the American Statistical Association},
  78(381):47--55.

\bibitem[{R Core Team}, 2024]{R2024}
{R Core Team} (2024).
\newblock {\em R: A Language and Environment for Statistical Computing}.
\newblock R Foundation for Statistical Computing, Vienna, Austria.

\bibitem[Spearman, 1904]{Spearman1904}
Spearman, C. (1904).
\newblock The proof and measurement of association between two things.
\newblock {\em The American Journal of Psychology}, 15(1):72--101.

\bibitem[Tango et~al., 2007]{Tango2007}
Tango, T.~M., Lichtman, M.~G., and Dolphin, A.~E. (2007).
\newblock {\em The Book: Playing the Percentages in Baseball}.
\newblock Potomac Books, Washington, DC.

\end{thebibliography}

\end{document}